# Ultra high energy resolution focusing monochromator for inelastic X-ray scattering spectrometer.

**Alexey Suvorov[*], Alessandro Cunsolo, Oleg Chubar, and Yong Q. Cai**

*National Synchrotron Light Source II, Brookhaven National Laboratory, Upton, NY 11973, USA*
[*]*asuvorov@bnl.gov*

**Abstract:** A further development of a focusing monochromator concept for X-ray energy resolution of 0.1 meV and below is presented. Theoretical analysis of several optical layouts based on this concept was supported by numerical simulations performed in the "Synchrotron Radiation Workshop" software package using the physical-optics approach and careful modeling of partially-coherent synchrotron (undulator) radiation. Along with the energy resolution, the spectral shape of the energy resolution function was investigated. It was shown that under certain conditions the decay of the resolution function tails can be faster than that of the Gaussian function.

## 1. Introduction

Improvement of the energy resolution in Inelastic X-ray Scattering (IXS) experiments down to 0.1 meV in a medium X-ray energy range (~10 keV) is a highly desired goal at many synchrotron radiation facilities[1-5]. This will allow expansion of the energy-momentum phase space boundaries accessible by the IXS into an uncharted regime precluded from various existing spectroscopy techniques. X-ray monochromator and analyzer are the two key constituents of any IXS spectrometer. In addition to the high energy resolution, it is desirable to have the monochromator with high spectral efficiency, high spectral contrast, and large range of energy tunability. Usually, the design of the high resolution monochromators contains multiple crystals arranged in a specific order and exploits highly asymmetric Bragg reflections at extreme grazing incidence. There are several factors which hamper the desired features to be attained. Firstly, the lattice parameter uniformity of the crystal optics must be better or at a level of $\Delta d/d \approx 10^{-8}$ which is at the verge of the best available quality of Si crystals. Secondly, the use of highly asymmetric Bragg diffraction with extreme grazing incidence makes the optics lengthy and imposes additional requirements on surface flatness and temperature uniformity. Moreover, the efficiency of the crystal optics deteriorates as the grazing angle approaches the critical angle of the surface specular reflection. Contrast of the spectral function is defined by the decay rate of its tails. The decay of the spectral tails in single-bounce crystal layouts is slow due to the fact that the crystal reflectivity tails follow the Lorentzian shape. In multi-bounce crystal layouts the reflectivity functions of the crystals are multiplied and the decay rate of the spectral tails can be improved. However, augmentation of the number of crystals in the monochromator impairs its efficiency and increases the complexity of the design. Application of anomalous transmission phenomenon in the monochromator design, like in [6], also can improve the decay rate of the spectral tails. However, the phenomenon introduces asymmetry to the spectral function shape which may complicate the subsequent spectral analysis of the experimental data.

Generally, the performance of crystal optics can be analyzed in the angular-energy phase space by means of DuMond diagrams [7]. Combination of focusing and crystal optics, referred to hereafter as focusing monochromator, introduces a new dimension, the real space distance, into the phase space [8, 9]. The principle of operation of the x-ray focusing monochromator is very similar to that of the Czerny-Turner monochromator [10] used in infrared, visible, and ultraviolet spectroscopies. The general idea behind it is to provide spatial separation of the focused beams according to their energies using angular dispersion of the diffracting optic. Thus the beam with the desired energy can be selected by an aperture in the image plane of the focusing optic. As we will demonstrate, the proposed focusing monochromator has the advantages of high spectral efficiency, high energy resolution, and wide energy resolution tunability, and allows potentially more compact design of the crystal optics.

In this paper we develop the theory of the x-ray focusing multi-crystal monochromator, and analyze the performance of several optical layouts. The results are supported by numerical simulations using partially coherent synchrotron radiation. Contrary to the matrix method of the geometrical optics presented in [9], our theory is based on the physical-optics formalism which gives a deeper insight into the processes involved. In particular, the presented theory allows the analysis of the spectral function and the decay rate of its tails in accordance with the optics parameters and configurations as used in synchrotron radiation beamlines. Most of the optics parameters for the numerical simulations were selected close or identical to the ones used at the NSLS-II IXS beamline. The obtained results provide a possible and practical pathway for a future monochromator upgrade of the beamline.

## 2. Basic theory

Generally x-ray diffraction from a crystal is described through reflectivity as a function of the transverse component of the incident wave vector in the plane of diffraction. Along with the paraxial approximation this allows factorization of the wave propagation integral into two independent terms. In this section we present a basic theoretical analysis of several optical layouts performed in the diffraction plane of the crystal optics. The transverse component of the incident wave vector normal to the diffraction plane of the crystal optics is a second order effect and is omitted in our analysis.

*2.1 Four-bounce HRM*

There exists a variety of designs and realizations of high resolution monochromators. In the following analysis we will consider a generic four-bounce in-line monochromator schematically shown in Fig.1. The outgoing beam of the in-line monochromator is coaxial with the incoming one. This is very convenient for synchrotron facilities since it keeps the x-ray beam along a straight line over large distances. The four-bounce high resolution monochromator (HRM) contains four asymmetrically cut crystals arranged in the so-called (+,-,-,+) diffraction geometry.

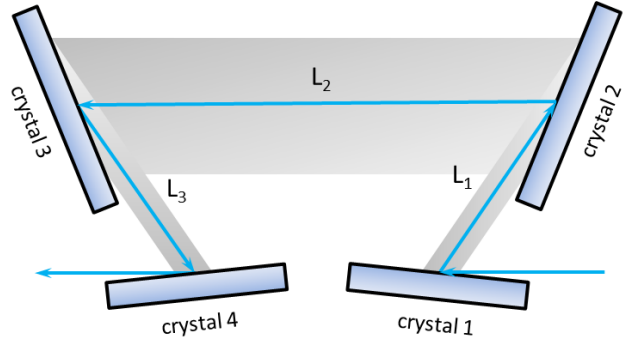

Fig. 1. Schematic of the HRM. $L_n$, $n$=1-3, denotes distances from the corresponding crystals.

For convenience, the coordinate system, directions of incident and diffracted waves, and angles of incidence and diffraction for the crystal 1 are shown schematically in Fig.2. After each crystal reflection the coordinate system of the diffracted beam can be obtained by rotating the coordinate system of the incident beam by the angle $\theta_{ni} + \theta_{no}$, $n$=1-4. The rotation angle has a positive sign if the rotation occurs in clockwise direction. By this means the signs of the rotation angles are arranged in the same manner as the diffraction geometry notation, i.e. (+,-,-,+). The origins of the coordinate systems for the incident and diffracted waves are located on the crystal surface, but for clarity, the systems in Fig.2 are annotated at the corresponding beam axes. The ($x$, $z$) plane is coplanar with the crystal diffraction plane and the $z$ axis is always directed along the beam propagation.

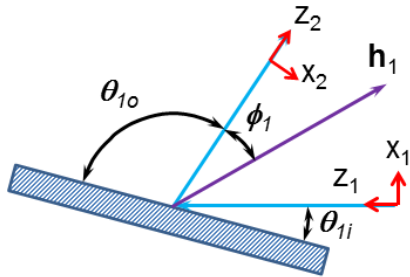

Fig.2. Diffraction geometry of the first crystal of the HRM. The angles of incidence and diffraction are denoted as $\theta_{1i}$ and $\theta_{1o}$, respectively. The angle of asymmetry $\phi_1$, is counted positively if $\theta_{1i} > \theta_{1o}$. For the purpose of clearness the origins of the local coordinate systems are annotated away from the crystal surface at the corresponding beam axes.

Consider a quasi-monochromatic scalar wave with the central energy of the spectral line $E_0$ ($E_0 = hc/\lambda_0$) and wave length $\lambda_0$ propagating from the right to the left in the diffraction plane of the crystals, and denote the relative deviation from the central energy as $\delta_E = (E-E_0)/E_0$. A solution for a plane wave diffraction by an ideal semi-infinite crystal is well known and can be expressed in terms of the Fourier transform integral [11]. In the case of the first crystal diffraction geometry (Fig. 2), the wave amplitude at a distance $L_1$ from the crystal can be written out as:

$$U_1(x_2) = \int \hat{U}_0(k_{0x}) R_1(k_{0x}) \exp\left(ik_{1x}x_2 - iL_1 \frac{k_{1x}^2}{2k}\right) \frac{dk_{0x}}{\sqrt{2\pi}}, \qquad (1)$$

where $\hat{U}_0(k_{0x})$ is the Fourier transform of the incident wave amplitude at the first crystal position $U_0(x_1)$, $R_1(k_{0x})$ is the complex reflectivity of the first crystal, and $k = 2\pi/\lambda$. A constant phase term in Eq.(1) was omitted. The transverse components $k_{0x}$ and $k_{1x}$ of the incident and diffracted wave vectors in their respective coordinate systems are related to each other through the refraction condition at the crystal boundary and can be written out as [8]:

$$k_{1x} = -b_1 k_{0x} + K(1+b_1)(\varphi_1 + \delta_E \tan\phi_1), \qquad (2)$$

where $b_1 = \sin\theta_{1i}/\sin\theta_{1o}$ is the absolute value of the crystal asymmetry factor, $\varphi_1$ is a small deviation of the crystal angular position from the maximum reflectivity, $\phi_1$ is the angle of asymmetry, and $K = 2\pi/\lambda_0$. The deviation angle $\varphi_1$ is counted positively if the angle of incidence is increased. The asymmetry angle $\phi_1$ is counted positively if $\theta_{1i} > \theta_{1o}$.

Applying the above formalism to the remaining crystals, one can derive the wave amplitude at a distance $L_4$ after the last crystal as:

$$U_4(x) = \int \hat{U}_0(k_{0x}) R_\Sigma(k_{0x}) \exp\left(ik_{4x}x - i\sum_{n=1}^{4} L_n \frac{k_{nx}^2}{2k}\right) \frac{dk_{0x}}{\sqrt{2\pi}},$$

$$R_\Sigma(k_{0x}) = \prod_{n=1}^{4} R_n(k_{(n-1)x}) \qquad (3)$$

The wave vector components relates to each other as:

$$k_{(n+1)x} = -b_n k_{nx} + \text{sign}(c_n) K(1+b_n)\Delta_n,$$
$$\Delta_n = \varphi_n + \delta_E \tan\phi_n \qquad (4)$$

where $c_n = (+1, -1, -1, +1)$ reflects the crystal arrangement in the layout. The distances $L_{1,2,3}$ between the crystals are usually much smaller than the distances upstream and downstream from the HRM. In addition, the beam propagating in-between the crystals is generally well collimated. Taking this into account, we can neglect the second order terms with $L_{1,2,3}$ in the Eq.(3) and rewrite it as

$$U_4(x) = \frac{1}{b_\Sigma} \int \hat{U}_0(k_{4x}) R_\Sigma(k_{4x}) \exp\left(ik_{4x}x - iL_4 \frac{k_{4x}^2}{2k}\right) \frac{dk_{4x}}{\sqrt{2\pi}} \qquad (5)$$

where $b_\Sigma = b_1 b_2 b_3 b_4$, $k_{4x} = b_\Sigma k_{0x} + K\alpha_\Delta$, and $\alpha_\Delta$ in the explicit form is

$$\alpha_\Delta = (1+b_4)\Delta_4 + b_4(1+b_3)\Delta_3 - b_3 b_4(1+b_2)\Delta_2 - b_2 b_3 b_4(1+b_1)\Delta_1 \qquad (6)$$

The parameter $\alpha_\Delta$ denotes angular deviation of the transmitted beam from the optical axis, i.e. from the z-axis of the corresponding coordinate system.

Generally, to keep the transmitted beam through the HRM coaxial with the incident one, all of the crystals in the monochromator should have the same reflection. The energy of the transmitted beam can be changed by rotating the first two and the last two crystals of the HRM in opposite directions. Substituting the angular deviations $\varphi_1 = \varphi_3 = \varphi$ and $\varphi_2 = \varphi_4 = -\varphi$ into the Eq.(6), we arrive at a well-known relation of the HRM energy as a function of angular position of the crystals $\Delta E = -E_0 \varphi/\tan\theta_B$ where $\theta_B$ is the Bragg angle of the crystal reflection.

*2.2 Optical layout I*

In the first optical layout we consider, the crystal optic is located in-between the focusing optic and its image plane (Fig.3). To simplify the analysis, it is supposed that the focusing optic is an ideal thin lens.

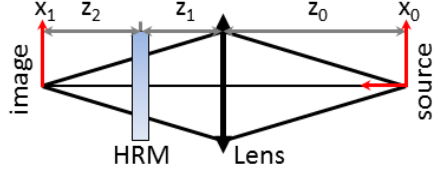

Fig.3. Schematic of the optical layout I.

We assume that the wave propagating through the system is Gaussian:

$$U_0(x) = \frac{1}{\sqrt{i\lambda q(z)}} \exp\left( ik \frac{(x-x_0)^2}{2q(z)} \right), \quad (7)$$

where $q(z) = z - iz_R$, $z$ is the distance from the source, $z_R$ is the Rayleigh range, and $x_0$ is the transverse coordinate of the source. Using the free-space propagation properties of the Gaussian beam, the wave amplitude at the HRM position can be presented as a Fourier integral:

$$U_0(x) = A_0 \int \exp\left\{ ik_{0x}\left(x + \frac{q_1}{q_0}x_0\right) + i(q_1 - z_1)\frac{k_{0x}^2}{2k}\right\} \frac{dk_{0x}}{\sqrt{2\pi}},$$

$$A_0 = \sqrt{-\frac{q_1}{q_0}} \exp\left( ik \frac{x_0^2}{2q_0}\left[1 + \frac{q_1}{q_0}\right]\right), \quad \frac{1}{q_1} = \frac{1}{f} - \frac{1}{q_0}, \quad q_0 = z_0 - iz_R \quad (8)$$

where $f$ is the focal distance of the lens and $k_{0x}$ is the x-component of the wave vector. Using Eq.(5), the wave amplitude at a distance $z_2$ from the HRM can be written as:

$$U_1(x_1) = \frac{A_0}{b_\Sigma} \exp\left(-i\frac{q_1}{b_\Sigma q_0}\alpha_\Delta x_0\right) \int R_\Sigma(k_{1x}) \exp\left\{ i(x_1 - \tilde{x}_1)k_{1x} - i(z_2 - z')\frac{k_{1x}^2}{2K}\right\} \frac{dk_{1x}}{\sqrt{2\pi}},$$

$$\tilde{x}_1 = \alpha_\Delta z' - \frac{q_1}{b_\Sigma q_0}x_0, \quad z' = \frac{1}{b_\Sigma^2}(q_1 - z_1) \quad (9)$$

The image plane distance, $z_2$, can be obtained from the condition $z_2 = \mathrm{Re}(z')$:

$$z_2 = \frac{1}{b_\Sigma^2}\left( \frac{z_0 f}{z_0 - f} - z_1 \right). \quad (10)$$

Equation (10) shows that the HRM scales the image plane distance of a single lens by a factor $b_\Sigma^2$. If the cumulative asymmetry factor of the HRM is $b_\Sigma \neq 1$, then the image plane distance can be significantly shorter or longer than that of the single lens system without the HRM. The intensity at the image plane can be represented as:

$$I_1(x_1) = \frac{M_1}{b_\Sigma} \left| \int R_\Sigma(k_{1x}) \exp\left\{ -i(\tilde{x}_1 - x_1)k_{1x} - z_R M_1^2 \frac{k_{1x}^2}{2K}\right\} \frac{dk_{1x}}{\sqrt{2\pi}} \right|^2,$$

$$M_1 = \frac{f}{b_\Sigma(z_0 - f)} \quad (11)$$

where the factor $M_1$ can be considered as the system magnification. Transverse position, $x_1$, of the focused beam in the image plane can be found from the condition $x_1 = \mathrm{Re}(\tilde{x}_1)$:

$$x_1 = \alpha_\Delta z_2 - M_1 x_0. \qquad (12)$$

Equation (12) shows that the system images the source with the magnification factor $M_1$ and shifts the image of the focused monochromatic beam by $\alpha_\Delta z_2$.

*2.3 Optical layout II*

In the second optical layout the multi-bounce crystal optic is located in-between the source and the focusing optic (Fig.4).

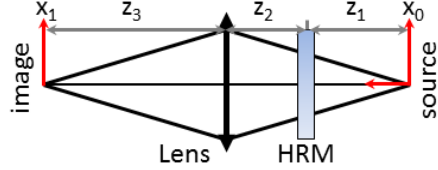

Fig. 4. Schematic of the optical layout II.

Using the same approach as in the previous section, it can be shown that the wave amplitude at a distance $z_3$ downstream from the lens is:

$$U_2(x_1) = \frac{A_1}{b_\Sigma} \int R_\Sigma\left(\frac{z_3}{z_s} k_{1x}\right) \exp\left\{i(\tilde{x}_2 - x_1)k_{1x} + i(z_s - z'')\frac{z_3^2}{z_s^2}\frac{k_{1x}^2}{2k}\right\} \frac{dk_{1x}}{\sqrt{2\pi}},$$

$$A_1 = \sqrt{-\frac{z_3}{z_s}} \exp\left\{ik\frac{x_1^2}{2z_i}\left(1 + \frac{z_s}{z_3}\right) + i\frac{1}{b_\Sigma}\alpha_\Delta x_0\right\}, \qquad (13)$$

$$\tilde{x}_2 = \frac{z_3}{b_\Sigma z_s}\left(\frac{1}{b_\Sigma}\alpha_\Delta q_1 - x_0\right), \quad z'' = \frac{q_1}{b_\Sigma^2} + z_2, \quad \frac{1}{z_s} = \frac{1}{f} - \frac{1}{z_3}$$

The image plane distance $z_3$ can be found from the condition $z_s = \operatorname{Re}(z'')$. It can be expressed as:

$$\frac{1}{z_3} = \frac{1}{f} - \frac{1}{z_0'}, \quad z_0' = \frac{z_1}{b_\Sigma^2} + z_2. \qquad (14)$$

The intensity in the image plane can be written as:

$$I_2(x_1) = \frac{M_2}{b_\Sigma}\left|\int R_\Sigma(b_\Sigma M_2 k_{1x}) \exp\left\{i(\tilde{x}_2 - x_1)k_{1x} - z_R M_2^2 \frac{k_{1x}^2}{2k}\right\} \frac{dk_{1x}}{\sqrt{2\pi}}\right|^2,$$

$$M_2 = \frac{f}{b_\Sigma(z_0' - f)} \qquad (15)$$

where the factor $M_2$ can be considered as the system magnification. The transverse position, $x_1$, of the focused beam in the image plane can be found from the condition $x_1 = \operatorname{Re}(\tilde{x}_2)$:

$$x_1 = M_2\left(\frac{1}{b_\Sigma}\alpha_\Delta z_1 - x_0\right). \qquad (16)$$

The obtained result can be interpreted as if the system creates a virtual source with a transverse coordinate $x_0' = x_0 - \alpha_\Delta z_1/b_\Sigma$ which is imaged with the magnification $M_2$.

*2.3 Optical layout III*

This layout is similar to the layout of the Czerny-Turner monochromator [10]. In this case the crystal optic is located in-between two lenses with the focal distances $f_1$ and $f_2$ (Fig. 5).

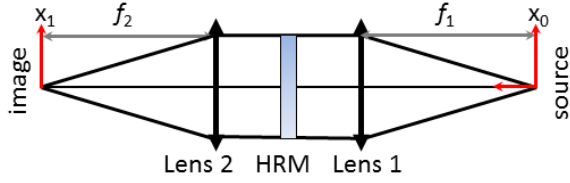

Fig. 5. Schematic of the optical layout III.

The first lens performs collimation of the incident radiation while the second one performs focusing. If we assume the distance between the lenses is reasonably small so that the associated second order term can be neglected, the intensity in the image plane can be written as:

$$I_3(x_1) = A \exp\left\{-\frac{2}{w_0^2 M_3^2}(\tilde{x}_3 - x_1)^2\right\}\left|R_\Sigma\left(k\frac{x_1}{f_2}\right)\right|^2, \quad (17)$$

$$A = 1/b_\Sigma M_3 \pi w_0^2, \quad \tilde{x}_3 = \alpha_\Delta f_2 - M_3 x_0, \quad M_3 = b_\Sigma \frac{f_2}{f_1}$$

where $w_0$ is the Gaussian beam waist size. In this case the intensity distribution in the image plane is proportional to the product of the Gaussian function of the focused beam with the reflectivity function of the HRM. The transverse position, $x_1$, of the focused beam in the image plane can be found from the condition $x_1 = \tilde{x}_3$:

$$x_1 = \alpha_\Delta f_2 - M_3 x_0. \quad (18)$$

This result is analogous to the optical layout I and can be interpreted as if the source is imaged with the magnification factor $M_3$ and is shifted by $\alpha_\Delta f_2$.

## 3. Energy resolution function

### 3.1 Resolution function width

The intrinsic energy resolution of the four-bounce HRM can be estimated using conventional DuMond diagram [7]. We note that the diffraction angle between the directions of the incident and reflected beams has a positive sign if it is measured clockwise. Then the counterclockwise direction of the diffraction angle is negative. To keep the beam transmitted through the HRM in-line with the incident one, a sum of all diffraction angles in the setup should be zero. If all of the crystal reflections have the same Bragg angle and the asymmetry factors of the first two crystals are reciprocal to the ones of the last two crystals, i.e. $b_4=1/b_1$, $b_3=1/b_2$, the intrinsic energy resolution of the four-bounce HRM for the sigma polarization of the incident beam can be approximately given by [1]:

$$\frac{\Delta E_{HRM}}{E_0} = \frac{\sqrt{b_1 b_2}\, \omega_s}{\tan\theta_B}, \quad (19)$$

where $\omega_s$ and $\theta_B$ are the Darwin width and the Bragg angle of the reflection, respectively.

The beam propagated through the focusing system and the HRM is deflected in the image plane due to the dispersion properties of the crystal optic and variations of crystal angular positions. If we assume that the crystals are fixed at the angular positions $\varphi_n = 0$, $n=1$-4, then, using Eqs. (4) and (6), the degree of the deflection can be expressed in terms of the dispersion rate [9]:

$$D = \frac{2}{E_0}\left[(1+b_4)\tan\phi_4 + b_4(1+b_3)\tan\phi_3 - b_3 b_4(1+b_2)\tan\phi_2 - b_2 b_3 b_4(1+b_1)\tan\phi_1\right], \quad (20)$$

which defines an angle, $\alpha_E = D\Delta E$, of the deflection of a monochromatic beam with energy $E = E_0 + \Delta E$ from the optical axis. The range of the angular dispersion is limited by the intrinsic energy resolution of the HRM, $|\alpha_E| \leq D\Delta E_{HRM}/2$. Because of this, the spatial range of the x-ray beam motion in the image plane is also limited. We will call this range as the dispersion range.

A slit placed in the image plane of the system will work as an energy filter. The intensity of the x-rays transmitted through the slit can be obtained by integrating the intensity in the image plane, Eqs. (11), (15), and (17), multiplied by a box function $u_a(x_1)$ of the slit which has non-zero values only within the slit aperture $-a/2 \leq x_1 \leq a/2$:

$$I_\Sigma(E) = \int I_n(\tilde{x}_n - x_1) u_a(x_1) dx_1, \quad (21)$$

where $n=1,2,3$ corresponds to the optical layout I, II, and III, respectively. Equation (21) represents energy resolution function of the system as a convolution of the image intensity with the slit box function.

Suppose that the transverse dimension of the slit is very small, equations (12), (16), and (18) demonstrate that the energy of the beam at the point $x_1=0$ depends on the coordinate of the x-ray point source. Let us consider a finite size x-ray source as a set of incoherent point sources. If the transverse size of the x-ray source is $S$, then the energy bandwidth of the beam downstream from the slit is

$$\Delta E_I = M_1 \frac{S}{Dz_2}, \quad \Delta E_{II} = b_\Sigma \frac{S}{Dz_1}, \quad \Delta E_{III} = M_3 \frac{S}{Df_2} \quad (22)$$

for the optical layouts I, II, and III, respectively. Estimates in Eq. (22) give the best energy resolution which can be achieved in the system with infinitely small slit. In all layouts the energy bandwidth gets smaller with the distances $z_{1,2}$ and $f_2$. In the first two layouts the smallest energy bandwidth is achieved when the crystal optic is located right next to the focusing optic. In the last layout, the exact position of the crystal optic with respect to the lenses is not significant under the given approximations.

*3.2 Resolution function shape*

Along with the energy bandwidth, the shape of the spectral line of the high energy resolution monochromator is an important characteristic of an x-ray spectrometer. Decay rate of the resolution function wings is a figure of merit of the spectrometer system. The sharper slopes of the resolution function improve contrast in the experimental energy-decay spectra. Commonly the decay rate of the resolution function of the crystal optics is compared to the Gaussian function. The metric of the decay rate can be established as a ratio of the function width at a level of 0.1% from its maximum to its width at half-maximum. Let us call this number as a decay factor. For the Gaussian function the decay factor, $\tau_G$, is about $\tau_G = 3.2$.

The shape of the resolution function in the focusing layout depends on three primary factors, namely the x-ray beam angular divergence/convergence, crystal optic angular acceptance, and the slit dimension at the image plane. Equations (11) and (15) demonstrate that the image intensity in the layouts I and II is proportional to the inverse Fourier transform of a product of two functions, namely the cumulative reflectivity function of the crystal optic and the Gaussian-shape intensity function of the propagating beam. The region of integration in these equations is determined by the smallest width of these functions. When the x-ray beam divergence / convergence is much larger than the angular acceptance of the HRM, the intensity in the image plane can be approximated by the Fourier transform of the HRM cumulative reflectivity function. It is well-known that the Fourier transform of the reflectivity function of a single crystal is related to the crystal propagator [11] or Green's function [12]. In case of single crystal diffraction it can be expressed as a combination of Bessel functions. The Fourier transform of the product of multiple reflectivity functions cannot be easily expressed in an analytical form. To illustrate this case we have performed numerical simulations which will be presented in the following section. When the x-ray beam divergence / convergence is much smaller than the angular acceptance of the HRM, the intensity in the image plane can be approximated by the Gaussian function.

The slit dimension affects not only the energy bandwidth of the transmitted beam but also its shape. There are at least three distinctive cases where significantly different resolution function shape may result. In the first case the slit dimension is smaller than the width of the intensity function in the image plane and tends to zero. Then the shape of the energy resolution function approaches the shape of the intensity function. According to the above analysis, when the x-ray beam is well collimated, the intensity function can be Gaussian and so is the energy resolution function. In the second case the slit dimension is much larger than the width of the intensity function but smaller than the dispersion range. As earlier, by dispersion range we assume the range of beam motion in the image plane due to dispersion effect and finite energy resolution of the crystal optic. In this case the energy resolution function approaches the box shape of the slit with the decay factor better than the Gaussian one. In the third case the slit is fully open and does not affect any of the beam properties. Then the energy resolution function follows the shape of the resolution function of the crystal optic. Commonly the resolution function of the crystal optics has the Lorentzian function wings with the decay factor larger than the Gaussian one.

## 4. Numerical simulations

In this section we will illustrate our theoretical analysis with the numerical examples. The simulations were performed using the "Synchrotron Radiation Workshop" (SRW) software package. The SRW is an open source package which was developed to provide accurate and efficient computation of synchrotron radiation and x-ray optics using wave-optics principles. The simulations were performed in the Department of Energy's National Energy Research Scientific Computing Center (NERSC).

We used an optical layout of the IXS 10ID beamline at the NSLS-II synchrotron facility [13] as the basis for the simulations. The IXS beamline will be operated primarily at photon energies near 9.13 keV. Its insertion device is a room-temperature 22-mm period In-Vacuum Undulator (IVU22) with a total length of 3 m installed in a high-β strait section of NSLS-II. The undulator was slightly detuned from the resonant photon energy to gain maximum transmitted intensity. In this case the x-ray wave field generated by the undulator is different from that of the Gaussian beam which results in a slightly different intensity shapes and wings. However, the principles of the focusing monochromator operation remains the same and the general theoretical estimates obtained earlier for the Gaussian beam are still valid.

The IXS beamline contains an assembly of four Beryllium compound refractive lenses (CRL), which were installed in the beamline front-end. Each CRL had an apex radius of 0.3 mm, apex thickness of 0.1 mm, and optical dimensions 2 mm by 1mm (HxV). The CRLS were located at a distance about 18.7 m from the undulator center. They were designed to provide a 1:1 focusing of the x-rays in the vertical plane. A secondary source aperture (SSA) was located in the image plane of the CRLs at a distance 37.4 m from the source.

The IXS four-bounce HRM uses the Si (642) reflections for each crystal. The design asymmetry angles of the crystals were $\phi_1=66°$, $\phi_2=65°$, $\phi_3=-65°$, and $\phi_4=-66°$. The corresponding asymmetry factors of the crystals were $b_4=1/b_1=12.2$ and $b_3=1/b_2=9.5$. According to Eq. (19), the theoretical energy resolution of the HRM is $\Delta E_{HRM}=1$ meV. The dispersion rate (Eq. (20)) of the HRM with the given parameters was $D=67.0$ μrad/meV.

In the first simulation the HRM was placed at a distance $z_2=3$ m upstream from the CRLs image plane according to the optical layout I. Because of the $b_\Sigma=1$, the magnification factor of the optical layout remains the same as for the single lens.

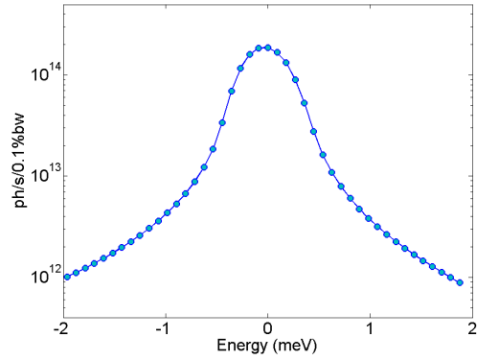

Fig. 6. Energy resolution curve simulated for the optical layout I with fully open slit. Energy FWHM is 0.6 meV.

Figure 6 demonstrates energy resolution curve of the layout in case of fully opened SSA. In this case the energy resolution width and shape are defined by the crystal optic only. The simulated energy full width at half maximum (FWHM) is $\Delta \tilde{E}_{HRM}=0.6$ meV. The discrepancy with the theoretical estimate comes from the fact that the numerical simulation takes into account incoming beam divergence and particular shape of the crystal reflectivity function while in the estimate the crystal reflectivity was approximated by a box-shape band. The wings of the energy curve follow the Lorentzian function and decay with a rate $\sim 1/x^2$. The decay factor of the simulated curve is $\tau_{HRM}=12.1$. The spectral efficiency of the HRM is about 37%.

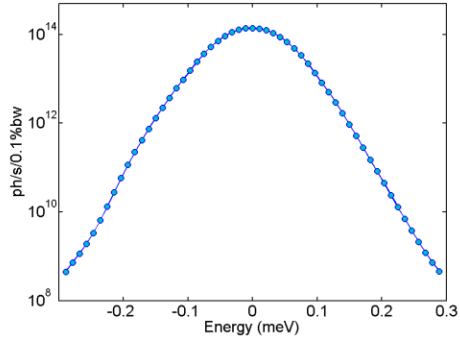

Fig. 7. Energy resolution curve simulated for the optical layout I with the vertical slit size of 17 μm. Energy FWHM is 0.1 meV

Figure 7 demonstrates energy resolution function of the same optical layout with the SSA closed down to 17 μm. The value of the slit opening was set equal to the FWHM of the monochromatic beam image in the SSA plane. The simulated energy curve has a FWHM $\Delta E_1$ =106.8 μeV with the decay factor $\tau_1$ =3.5 which is very close to the Gaussian one. The spectral efficiency of the layout, which takes into account the intensity losses due to the finite slit size, is about 73%.

To improve the resolution function decay rate and spectral efficiency, the beam incident on the crystal optic has to be collimated much better than the angular acceptance of the crystal optic. This requirement is fulfilled in the two-lens layout III shown in Fig. 5.

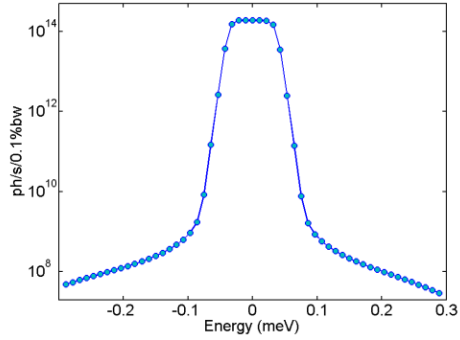

Fig. 8. Energy resolution curve simulated for the optical layout III with the vertical slit size of 25 μm. Energy FWHM is 0.07 meV.

Figure 8 demonstrates simulated energy curve of the optical layout III with the following parameters: $f_1$=18.7 m, $f_2$=5 m, and slit size $SSA$=25 μm. The first focusing element contained two CRLs and the second one contained 5 CRLs with parameters analogous to the previous case. With these parameters the magnification factor of the layout was $M_3$=0.27 and the vertical FWHM of the monochromatic beam image was about 4.6 μm in the SSA plane. Using Eq. (18), the energy resolution in the layout can be estimated as $\Delta E = SSA/Df_2$ =74.6 μeV which corresponds exactly to the FWHM of the simulated curve. Due to the fact that the SSA size is larger than the monochromatic beam image, the shape of the energy curve approaches the box function of the slit with the decay factor $\tau_2$ =1.7. The obtained decay factor is about two times better than that of the Gaussian function. The spectral efficiency of the layout is 92%.

### 5. Conclusion

Combination of dispersive crystal optics with focusing optics allows monochromatization of x-ray radiation to a level beyond 0.1 meV with a very high spectral efficiency. In fact, the high degree of x-rays monochromatization can be achieved with moderate crystal optics parameters in terms of operation energies and reflection orders. The lessened requirements on the asymmetry angles, Bragg angles and reflection orders of the crystal optics lead to a more compact and efficient design of the high energy resolution monochromators. Variation of certain distances in a synchrotron beamline layout may allow the realization of tunable energy resolution depending on particular experiment requirements.

The shape of the resolution function depends on the incident x-ray beam parameters, crystal optics, and transverse dimension of the slit aperture in the focusing optics image plane. It may vary from Lorentzian shape to the Gaussian and box-like ones with significantly different decay rates.

Single-lens optical layout has the advantage of high performance and simplicity. However, it does not allow large variation of the magnification factor because of the limited angular acceptance of the crystal optics. The two-lens optical layout allows more freedom in selecting crystal parameters and magnification factors. Also, in some cases it can lead to extremely high spectral efficiency and a spectral line shape close to the box function with high decay rate of the function wings.

Although the high energy resolution in the focusing layout can be achieved with moderate energy resolution of crystal optics, the requirements on the crystal lattice quality remains the same. Non-uniformity in the crystal lattice invalidates the dispersion effect and consequently the achievable energy resolution. Having a compact design of the crystal monochromator and smaller x-ray beam sizes provided by focusing optics therefore reduces this challenge significantly.

**Acknowledgments**

Work at Brookhaven was supported by the U.S. Department of Energy under contract No. DE-SC0012704. This research used resources of the National Research Scientific Computing Center, a DOE Office of Science User Facility supported by the Office of Science of the US Department of Energy under contract No. DE-AC02-05CH11231.